\documentclass[twocolumn,prl,showpacs,preprintnumbers,amsmath,amssymb]{revtex4}
\usepackage{graphicx}
\usepackage{dcolumn}
\usepackage{bm}

\begin{document}

\title{Tunneling and Non-Universality in Continuum Percolation Systems}
\author{C. Grimaldi$^{1,2}$}
\author{I. Balberg$^3$}
\affiliation{$^1$ LPM, Ecole Polytechnique F\'ed\'erale de
Lausanne, Station 17, CH-1015 Lausanne, Switzerland}
\affiliation{$^2$DPMC, Universit\'e de Gen\`{e}ve, 24 Quai
Ernest-Ansermet, CH-1211 Gen\`{e}ve 4, Switzerland}

\affiliation{$^3$The Racah Institute of Physics, The Hebrew
University, Jerusalem 91904, Israel }

\begin{abstract}
The values obtained experimentally for the conductivity critical
exponent in numerous percolation systems, in which the
interparticle conduction is by tunnelling, were found to be in
the range of $t_0$ and about $t_0+10$, where $t_0$ is the
universal conductivity exponent.  These latter values are however
considerably smaller than those predicted by the available ``one
dimensional"-like theory of tunneling-percolation.  In this letter
we show that this long-standing discrepancy can be resolved by
considering the more realistic ``three dimensional" model and the
limited proximity to the percolation threshold in all the many
available experimental studies
\end{abstract}

\maketitle

While the existence of non-universality in physical properties of
percolation systems has been established some eighteen years ago
\cite{feng,balb1} the discrepancy between the numerous
experimental results and the corresponding available theories is
still an unresolved issue \cite{balb2,balb3}.  In particular the
values of the critical exponent of the electrical conductivity,
$t$, reported in hundreds of works
\cite{balb3,vionnet,note,ali,lee,heaney} on various composite
materials in the last twenty years, are not quantitatively
accounted for by those theories.  The major difficulty in the
comparison of the theoretical predictions with the experimental
results appears to be the lack of the experimental
geometrical-structural information that is expected to yield the
diverging (but normalizable) distributions of the
local-conductances.  The latter is the well known
\cite{feng,balb1,balb2} pre-requisite for the non-universal
behavior \cite{kogut}. The essence of the non-universal behavior
is that the global resultant resistance of a percolation system
that is given by
\begin{equation}
\label{eq1} R_t\propto (p-p_c)^{-t}
\end{equation}
is determined by
\begin{equation}
\label{eq2} R_t\propto \langle R\rangle(p-p_c)^{-t_0}
\end{equation}
where $\langle R\rangle$ is the average value of the ``bond''
(inter-nearest-neighbour conducting particles) resistance in the
system, $p$ is the bond occupation probability, $p_c$ is the
threshold of the system's electrical connectivity, $t$ is the
conductivity exponent and $t_0$ is the ``universal" critical
exponent that is determined solely by this connectivity.  Since
in a random system there is no correlation between the geometrical
position of a bond and its resistance, any random subsystem of
$p_c$ bonds for a given $p$, will provide a connected-conducting
network.  In particular, if the latter network is chosen by the
descending values of the bond conductance, $g$, the value of
$\langle R\rangle$ will be an average that is determined by $g_c$,
the smallest value of $g$ in that sub-network \cite{kogut}.  In
the case where the distribution of the $g$ values, $h(g)$,
diverges as $g\rightarrow 0$, the value of $g_c$ will diminish as
$p$ approaches $p_c$, yielding a diverging behavior of $\langle
R\rangle$.  This behavior has been demonstrated by Kogut and
Straley \cite{kogut} (KS) for the distribution:
\begin{equation}
\label{eq3} (1-\alpha)g^{-\alpha}
\end{equation}
yielding that $\langle
R\rangle=[(1-\alpha)/\alpha](g_c^{-\alpha}-1)$. Hence, for the
non-diverging case ($\alpha < 0$), $\langle R\rangle$  is finite
while, for ($0 < \alpha < 1$) the diverging (but normalizable)
case, one finds that \cite{kogut}: $\langle R\rangle\propto
(p-p_c)^{-t_n}$ where, for the distribution given in Eq.
(\ref{eq3}), $t_n=\alpha/(1-\alpha)$. Hence the non-universal
contribution to the conductivity exponent ($t_n = t-t_0$) is
determined only by $\langle R\rangle$ and thus we are able to
limit our discussion here to the evaluation of this quantity.

Turning to the problem at hand, \textit{i.e.} the large
quantitative discrepancy between the theoretical predictions and
the experimental observations, we note that in both systems for
which theories were advanced, the porous media \cite{feng} and the
tunneling percolation problem \cite{balb1} (see below), the
conductors distribution was mapped onto the KS \cite{kogut}
distribution (Eq. (\ref{eq3})) yielding specific predictions as to
the values of $t_n$. However, the experimental results were found
to be, in general, larger than expected in the first class of
systems \cite{balb2} and smaller or much smaller
\cite{balb1,balb3,vionnet,note,ali,lee,heaney} than expected in
the second class of systems \cite{balb1,balb3,rubin}. One of us
has explained \cite{balb2} the first observation by applying
phenomenologically a modified $h(g)$ distribution that still
retains the KS dependence (Eq. (\ref{eq3})). That approach yielded
new limits to the theoretical predictions that seem to accommodate
all relevant experimental data \cite{balb2}.

In this letter we seek a general understanding of the
non-universal behavior of the second class of systems,
\textit{i.e.} in systems in which the conducting particles are
embedded in an insulating matrix and the transport between the
particles is by tunneling \cite{balb1,balb3,vionnet,rubin}. While
some qualitative explanations \cite{balb3,heaney,rubin,radha} were
proposed for the above mentioned discrepancy in those systems,
they were unable to account systematically for it
\cite{heaney,planes}. Moreover, only qualitative specific
explanations have been given \cite{chiteme,breeze}  to the fact
that in some conducting composites the critical behavior of the
transport has been found \cite{note,chen,mandal,chiteme} to be
composed of a few ``non-universal" regimes.  We note in passing
here that the dominant network that contributes to $R_t$ is that
of the nearest neighbors' network that yields in practice a bona
fide percolation system \cite{balb3,toker}. This is of course
because of the exponential decay of the tunneling probability
with the interparticle distance $r$. Correspondingly, the possible
divergence of the average ``bond'' resistance $\langle R\rangle$
is determined by the largest $r$'s of the nearest neighbours. The
corresponding distribution of the nearest neighbours distance in
the continuum, $P(r)$, that was considered previously
\cite{balb1,rubin} in order to evaluate the origin of the
nonuniversality of the tunnelling-percolation model within the
framework of the KS model, was the ``one dimensional" (1D) Hertz
distribution \cite{efros,torqua}:
\begin{equation}
\label{hertz} \frac{1}{a-b}\exp\left(-\frac{r-b}{a-b}\right).
\end{equation}
Here $r$ is the distance between the centers of two nearest
neighbor spherical conducting particle, $b$ is their diameter and
$a$ is the average nearest-neighbour interparticle (three
dimensional) distance that can be estimated from
\cite{balb1,balb2,balb3}: $\frac{4\pi}{3}(a/2)^3N=1$   where $N$
is the density of the conducting particles. Combining this 1D
distribution with the simple (normalized, $0 \leq g \leq 1$)
tunneling dependence of the interparticle conductance on $r$
\cite{balb1,balb3,rubin}, $g(r)=\exp[-(r-b)/d]$, where $d$ is the
typical tunneling decay parameter, one obtains an $h(g)$
distribution of the form given in Eq. (\ref{eq3}) but with
\cite{balb1,balb3,rubin}:
\begin{equation}
\label{alpha} \alpha=1-\frac{d}{a-b}.
\end{equation}

While this 1D-like theory, leading to the simple analytic
prediction of Eq.(\ref{alpha}), appeared to capture the essence
of the problem and yielded the very simple and convenient physical
parameter $d/(a-b)$ for the description of the physical system, it
yields $t$ values that were much larger than the very many values
of $t$ found experimentally.  For example, the $(a-b)/d$ values
in many composites \cite{vionnet,note,ali,lee,heaney} are
expected \cite{balb3,note,rubin}, according to the 1D-like
phenomenological model, to be of the order of $50$ ($d \approx 1$
nm, $a-b \approx 50$ nm) but the highest observed
\cite{vionnet,ali} values of $t$ in corresponding composites was
about $10$. Another related phenomenon, noticed by us, following
the compilation of many experimental data \cite{balb3,vionnet},
is the general trend of the decrease of $t$ with the increase of
the critical volume fraction of the conducting phase, $v_c$, as
well as the scatter of the observed $t$ values within the $t_0$
and (about) $t_0+10$ interval for a priori similar systems
\cite{note,lee,rubin}. On the theoretical end, while the above
mentioned simplified model was, for the sake of simplicity
\cite{balb1}, a 1D-like model, no trials were made to test the
consequences of this 1D-like simplification, both conceptually
and mathematically, and the findings of one of us \cite{grima}
concerning the dependence of $h(g)$ on the dimensionality has not
been translated to a prediction of $t$.

In the present letter we show that the consideration of the
effect of dimensionality beyond the 1D simplification yields even
a qualitatively different behavior. In particular, the
corresponding new predictions of the higher dimension model
enables to account for the above-mentioned collection of
experimentally observed phenomena.  In fact, we have realized
already \cite{balb1} upon the introduction of the 1D-like model
that the probability of finding a nearest neighbor should
decrease slower than the decrease of $g$ with increasing $r$ in
order to enable a diverging distribution of $h(g)$.  This is
since if this is not the case (\textit{i.e.} $a < d$, or $\alpha
< 0$) a non-diverging $h(g)$ and thus a universal behaviour of
$R_t$ will be obtained.  In the 3D case the leading term in the
$P(r)$ distribution \cite{efros,torqua} is of the form of
$\exp[-(r^3-b^3)/(a-b)^3]$ (see below) and correspondingly, for
any values of $(a-b)$ and $d$, for large enough $r$, $P(r)$
decreases faster than $g(r)$ yielding a non-diverging $h(g)$ as
$g\rightarrow 0$, and thus, the asymptotic $p\rightarrow p_c$
critical behaviour is expected to be universal. On the other hand
this situation cannot be mapped onto a simple KS-like result for
$t_n$ and another framework is needed in order to evaluate the
$t$ values that are to be compared with the experimental
observations.  However, since $\alpha$ encloses the physical
information of the tunneling-percolation system we keep using it
for the system characterization in all dimensions.

The approach we have chosen is the effective medium approximation
(EMA) \cite{kirk}. This choice is justified following its
validity in general \cite{luck} and for the determination of
$\langle R\rangle$ in composite systems \cite{bergman} in
particular, as well as our above realization that the
contribution $t_n$ comes only from the average $\langle
R\rangle$.  Hence, the fact that the universal conductivity EMA
exponent, $t_0$, is $1$, and that $p_c$ is $1/3$ in the cubic
lattice, rather than the values expected from percolation theory
($\sim 2$ and $\sim 0.25$ respectively), simply means that the
$t_0$ value acts as a reference for the $t_n$ value that we are
examining (see below). This yields that, considering the
$t_0\approx 2$ value for the universal 3D percolation system
\cite{feng,balb1,balb2,balb3}, the ``correct" value of $t$ will be
larger than the $t$ value derived from our EMA calculations just
by a unity.

\begin{figure}[t]
\protect
\includegraphics[scale=0.38]{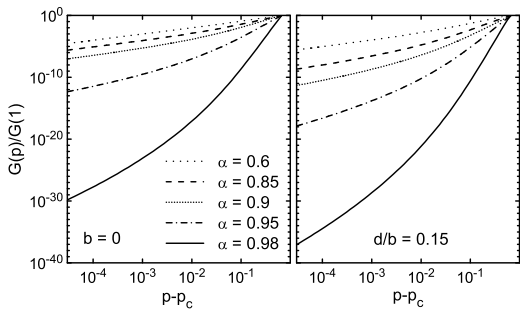}
\caption{Left panel: Our calculated EMA conductance $G$ as a
function of $p-p_c$ for ``dot'' particles ($b = 0$) with a nearest
neighbour distribution that is given by Eq.(\ref{torqua}). The
different cases refer to different values of the parameter
$\alpha = 1-d/(a-b)$. Right panel: the same as for the left panel
but for hard-core spheres with diameter $b > 0$. The cases shown
have the same $\alpha$ values as those in the left panel.}
\label{fig1}
\end{figure}

\begin{figure}[t]
\protect
\includegraphics[scale=0.38]{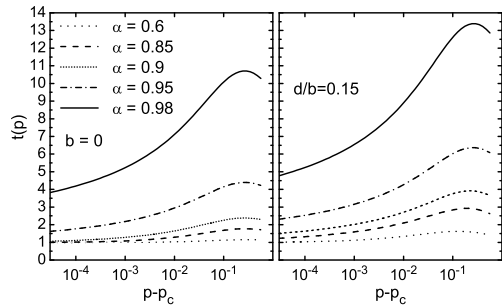}
\caption{The ``local'' transport exponent $t(p)$,
Eq.(\ref{tlocal}), as derived from the calculated data shown in
Fig. \ref{fig1}.} \label{fig2}
\end{figure}

Turning to the EMA calculation we consider a bond percolation
model for a cubic lattice with a bond conductance distribution
function of the form: $\rho(g)=ph(g)+(1-p)\delta(g)$  where the
non-zero conductance values are distributed according to $h(g)$.
Assuming that the nearest-neighbour inter-particle distances $r$
are distributed according to a given distribution function $P(r)$,
$h(g)$ reduces to: $h(g)=\int_0^\infty \!dr\,P(r)\delta[g-g(r)]$,
yielding \cite{kogut} the EMA average bond conductance $G$ as the
solution of the following integral equation:
\begin{equation}
\label{ema1}
\int_0^\infty\!dr\,\frac{P(r)}{g(r)+2G}=\frac{p-p_c}{2Gp}.
\end{equation}
For a 3D homogeneous dispersion of impenetrable spheres of
diameter $b$, $P(r)$ is well approximated by \cite{torqua}:
\begin{eqnarray}
\label{torqua} P(r)&=&\frac{24v(\gamma_1x^2+\gamma_2
x+\gamma_3)}{b}\exp\left[-8v\gamma_1(x^3-1)\right.\nonumber \\
&&\left. -12v\gamma_2(x^2-1)-24v\gamma_3(x-1)\right] \theta(x-1),
\end{eqnarray}
where $x = r/b$, $0 < v  <1$ is a dimensionless parameter
(coinciding with the volume fraction of the conducting
inclusions), $\theta$ is the unit step function and
\begin{equation}
\label{para}
\gamma_1=\frac{1+v}{(1-v)^3},\,\gamma_2=-\frac{v}{2}\frac{3+v}{(1-v)^3},\,
\gamma_3=\frac{1}{2}\frac{v^2}{(1-v)^3}.
\end{equation}
The parameter $v$ controls the value of the mean nearest-neighbor
interparticle distance $a$ (that we used in solving Eq.
(\ref{ema1})) through the relation: $a=\int_0^\infty\!dr\, rP(r)$.
From Eqs. (\ref{ema1}) and (\ref{torqua}) it is clear that the
effective medium conductance $G$ is governed by the parameter
$a/b$ that characterizes $P(r)$ and by the tunneling parameter
$d$.  A numerical solution of the integral equation given in Eq.
(\ref{ema1}) is plotted in Fig. \ref{fig1} for the case of
``dot'' particles ($b = 0$, left panel) and for hard-core spheres
with the typical \cite{balb3} $d/b = 0.15$ value (right panel).
The different plots of $G$ are shown for different values of the
characteristic parameter $\alpha$ (see Eq.(\ref{alpha})). In
order to better appreciate the behaviour of $G$ as $p\rightarrow
p_c$, the ``local" transport exponent defined as
\begin{equation}
\label{tlocal} t(p)=\frac{d\ln(G)}{d\ln(p-p_c)}
\end{equation}
is plotted in Fig. \ref{fig2} for the data exhibited in Fig.
\ref{fig1}. It is clear that for small values of $\alpha$ the
local exponent is only weakly dependent on $p$ and it is very
close to $t = 1$, \textit{i.e.} to the universal value of the
EMA.  For larger $\alpha$ values, $t(p)$ acquires a stronger
$p-p_c$ dependence which would correspond to an apparent
non-universality when $p$ is not too close to $p_c$. However, as
$p\rightarrow p_c$, the local exponent asymptotically reduces to
the universal value of $t_0 = 1$.

We have seen that our predicted $t$ values depend on the
proximity, $p-p_c$, to the percolation threshold $p_c$ and thus,
as we turn to the discussion of the experimental observations, we
have to compare the parameters that are commonly used to quantify
this proximity in the continuum with the above, lattice, $p-p_c$
parameter.  We note in particular that in the latter case $p_c$ is
of the order of unity ($0.247$ in bond percolation, or $1/3$ in
EMA, on the cubic lattice).  On the other hand in the continuum
one commonly \cite{vionnet,note} considers the fractional volume
(weight) content of the conducting phase, $v$ ($w$), and its
critical value $v_c$ ($w_c$). However, the latter critical values
can be vanishingly small \cite{balb2,vionnet,chiteme,pike} and
thus, while the absolute values of $v-v_c$ (or $w-w_c$) may be
very small compared to unity they do not correspond to a close
proximity to $v_c$ or $w_c$.  Hence, in general, the proximity in
both cases is better described \cite{celzard} in the present
context by $(p-p_c)/p_c$ and $(v-v_c)/v_c$ or $(w-w_c)/w_c$.
Since we noted that $(p-p_c)/p_c \sim (p-p_c)$, the comparison of
the $t$ values obtained in the theory and the experiment has to
be made for the same values of $p-p_c$ and $(v-v_c)/v_c$ or
$(w-w_c)/w_c$. Moreover, to appreciate the experimental
``resolution" limits of $(v-v_c)$ or $(w-w_c)$ that are achievable
thus far in composites, in general, and in systems in which the
percolation-tunneling model applies in particular, let us
consider the co-sputtered granular metals
\cite{balb3,toker,abeles}. For these systems one can typically
achieve a $50$ fold division of the sample with values of $v$
that vary from (ideally) $0$ to $1$. Hence, for the typical
\cite{abeles} $v_c \sim 0.5$ the smallest $v-v_c$ interval that
can be examined, away from $v_c$, is $0.02v_c$, yielding that the
closest proximity of $p-p_c$ ($p_c \sim 0.25$, see above) is not
smaller than about $0.01$.  As far as we know this is about the
closest proximity achieved thus far in the study of experimental
systems and thus all the available experimental data correspond
to the lattice percolation range of, at most, $1 \gtrsim p-p_c
\gtrsim 0.01$.

Examining our above EMA results (Fig. \ref{fig2}) in light of the
above considerations we see that over the above widest
experimentally achievable $p-p_c$ range the value of $t$ may be
taken as a constant with a deviation of not more than $\pm 20$
\%. Hence, it is not surprising that the common fit done in the
literature for experimental results is taken as representing a
single $t$ value over the accessible (less than two orders of
magnitude) $p-p_c$ range, while in fact a variable $t$ is present
over that range. Indeed, as we noted above, indications for the
variation of the measured transport exponents over the above
$p-p_c$ range can be found in the literature
\cite{note,chen,mandal,chiteme}. Also, the fact that the
``measured'' $t$ values are scattered, and vary between very
similar composites \cite{note,lee,chiteme}, but within the limits
of $t_0$ and $t$ of the order of $t_0+10$, as suggested here,
indicates that these observations follow a combination of the
small variation in internal system parameters ($a$, $b$, $d$) and
the limited $p-p_c$ intervals that are considered. On the other
hand, if higher experimental resolutions will be achieved (in the
preparation of series of samples) a more detailed verification of
our present EMA predictions, concerning the $p-p_c$ dependence of
$t$, is expected to be realized.

As we saw in Fig. \ref{fig2} the peak in the $t$ values shifts to
smaller $p-p_c$ values with the increase of $(a-b)/d$,
\textit{i.e.} higher $t$ values will be observed the larger the
value of $(a-b)/d$ (or $\alpha$), for the accessible $p-p_c$
range.  For this range this is qualitatively similar to the
behavior to be expected from the 1D-like model, but it is by far
more moderate in this range. However, the most important finding
is that the $t$ values predicted here are of the order of those
observed experimentally for the $a$, $b$ and $d$ parameters that
characterize the studied composites.  In fact our present
findings that yield relatively low (compared to the 1D-like
prediction $t_0+(a-b)/d-1$) $t$ values confirm our above
conjecture that it is the, diverging 1D and the non-diverging 3D,
$¨h(g)$ distributions that are responsible for the different
behavior of the 1D-like and the higher-dimension
percolation-tunneling systems.  The other general trend of the
many experimental results is that the smaller the $v_c$, (or
$w_c$) values (diminishing \cite{grima} even below 0.01) the
larger the $t$ value. This is well explained now by the above
mentioned fact that the $v-v_c$ values in these composites are
much smaller than the (proper) lattice-$p-p_c$ values, and thus
their farther proximity to the threshold in these composites,
yields, as seen in Fig. \ref{fig2}, larger $t$ values.

In conclusion, we have shown that the values expected to be
measured for the conductivity exponent, $t$, in
tunneling-percolation systems in the continuum, are between $2$
and the order of $10$ for typical ratios of the tunneling decay
constant and the size of the conducting particles.  The
dependence of $t$ on the proximity to the percolation threshold
accounts for the many reported experimental values and for their
scatter between the above values, as well as for the increase of
the $t$ values with the diminishing percolation threshold when it
is characterized, as is usually the case, by the fractional
volume or fractional weight of the conducting phase in the system.

Acknowledgement: This work was supported in part by the Israel
Science Foundation.


\begin{thebibliography}{99}
\bibitem{note}
A more detailed compilation of experimental observations and
their comparison with previous and present theoretical
predictions will be given elsewhere.
\end{thebibliography}
\end{document}